\documentclass[fleqn,12pt]{wlscirep}
\usepackage{amsmath}
\usepackage{graphicx}
\usepackage{footmisc}
\usepackage{color}

\begin{document}
\title {Mean field approach plus the Bethe ansatz solution  of one
dimensional models of Kondo insulator} 
\author[1]{Igor N. Karnaukhov} 
\affil[1]{G.V. Kurdyumov Institute for Metal Physics, 36 Vernadsky Boulevard, 03142 Kyiv, Ukraine}
\affil[*]{karnaui@yahoo.com}
\begin{abstract}
The Kondo insulator (KI)  is studyed in the frameworrk of one dimensional models of Kondo  and symmentic Anderson lattices at half filling.
The consistent application of the mean field approach and the solution using the Bethe ansatz made it possible to come close to solving the problem of the ground state of KI.  It is shown, that in  ${Z}_2$-field  electrons and local moments form  singlets at each lattice site.
This field leads to an on-site   interaction, on which electrons are scattered.  In the Kondo  chain
the Kondo insulator is similar to the Mott insulator in the Hubbard chain. The advantage of proposed approach is that it is possible to solve the problem for arbitrary  exchange interaction  constant.
In the Anderson lattice, the indirect on-site interaction between band electrons is attractive, which makes it possible to propose a new mechanism for high $T_c$ superconductivity in real compounds.

\end{abstract}

\maketitle

\section*{Introduction}

The Kondo and Anderson  lattice models decribe co-existence of interacting delocalized and localized states in crystals, formation of KI at half filling.   In contrast to the Anderson chain, where $d$-electrons form  flat bands, in the Kondo chain,  the localized electrons  are represented by magnetic moments regularly located at lattice sites. 
The exact solutions of the Kondo problem and the Anderson model made it possible to study the behavior of the electron liquid in the case of a weak interaction in a continuum approach \cite{1,2}. In the weak interaction limit, the Kondo temperature (the corresponding energy scale) is exponentially small in  the antiferromagnetic exchange interaction constant.

It is well known that the scattering of electrons by a local moment with spin flip dominates and forms the singlet state of the electrons and local moment in the Kondo problem. Most likely it is this scattering that also forms KI so it must be explicitly taken into account in the calculations.
Despite the rather great scientific interest in KI problem \cite{K1,K2,K3,K4,f,f1} we do not even know which ground 
 state corresponds to this phase. It is impossible to say unambiguously about the behavior of the electron liquid in KI even in the case of a weak interaction, since the approach to solving the problem is non-perturbative.  At the same time, a sufficiently large gap in the spectrum of KI 700 K in $FeSi$ \cite {f1} cannot be explained in  a weak coupling limit.

The static  ${Z}_2$-field is introduced in the mean field approach. This field leads to the formation of
 singlets (electron and local moment at the same lattice site) and an on-site repulsive interaction of band electrons.
 The scattering matrix of electrons has the form as in the Hubbard chain, so this scattering of electrons can be taken into account exactly. This makes it possible to solve KI problem in the Kondo chain without being limited to the weak interaction.
As one would expect,  in the symmetric  Anderson chain  with the integer valency of $d$-states, KI is similar to that in the Kondo chain. In contrast to the Kondo lattice, in the Anderson lattice an inderect on-site interaction of band electrons is attractive. 
This unique behavior leads to the formation of a high-$T_c$ superconductivity phase, because in this case we are talking about 
the electronic mechanism of electron pairing in compounds with a complex electronic structure.

\section*{The Kondo chain}

We will start with the  Hamiltonian of the spin-$\frac{1}{2}$ Kondo chain ${\cal H}_{K}={\cal H}_0+{\cal H}_{exch}$ 
\begin{eqnarray}
 &&{\cal H}_0= - \sum_{<i,j>}\sum_{\sigma=\uparrow,\downarrow}
c^\dagger_{i \sigma} c_{j \sigma}
, \nonumber \\ 
&&{\cal H}_{exch}= J\sum_{j=1}^N (s^+_{j}S^-_j+s^-_{j}S^+_j), \end{eqnarray} 
where 
 $c^\dagger_{j \sigma}$ and $c_{j \sigma}$ are the fermion operators  for an electron determined on a lattice site $j$  with spin $\sigma =\uparrow,\downarrow$, the hopping integral between the nearest-neighbor lattice sites is equal to one, the spin operators for the electron at
the site $j$  are determined as  $s^+_{j}=c^\dagger_{j \uparrow} c_{j \downarrow}$, $s^-_{j}=c^\dagger_{j \downarrow} c_{j \uparrow}$,   $\textbf{S}_j$ is the spin-$\frac{1}{2}$ operator determined on the lattice site $j$, $J\geq 0$ is the magnitude of an anisotropic antiferromagnetic exchange  interaction,  N is the total number of lattice sites.  

The scattering of electrons with spin flip forms the ground state of the electron liquid in the Kondo problem and the Kondo lattice, so  we will take into account only the anisotropic exchange interaction in (1) and consider the behavior of an electron liquid at half filling \cite{K1,K2,K3,K4}.

\subsection*{Solution of the problem}

We use the  following presentation for the ${\cal H}_{exch}$ term (below we will follow  \cite{K4} up to formula (4)):
\begin{eqnarray} 
&&J (s^+_{j}S^-_j + s^-_{j}S^+_j)=-J (s^-_{j}-S^-_j)^\dagger(s^-_{j}-S^-_j) -(m_{j}^2-m_{j}).
\end{eqnarray}
Using the Hubbard-Stratonovich transformation, we introduce a static ${Z}_2$- field  in a mean field approach 
$
  \lambda_{j}(c^+_{j\uparrow}c_{j \downarrow}-S^+_{j})+\lambda^*_{j}(c^+_{j\downarrow}c_{j\uparrow}- S^-_{j} )+2\mu_{j} m_{j}$. here   $m_{j \sigma}=c^\dagger_{j \sigma} c_{j, \sigma}$, $m_{j}=m_{j \uparrow}+m_{j \downarrow}$ are the density operators.

The effective Hamiltonian  is determined by an one-component ${Z}_2$-field, the $\mu$-component determines the Femi level.  In the model Hamiltonian (1)  and effective Hamiltonian the total number of electrons is conserved,  we will not take into account the  $\mu$-component. 

The effective Hamiltonian ${\cal H}_{eff}$ has the following form
\begin{eqnarray} 
&&{\cal H}_{eff}={\cal H}_0+
\sum_{j}[\lambda_{j}(c^+_{j \uparrow}c_{j \downarrow}-S^+_{j})+\lambda^*_{j}(c^+_{j \downarrow}c_{j \uparrow}-S^-_{j})] +\sum_j \frac{\lambda_j^2}{J}.
\end{eqnarray} 

The solution of an  one-particle wave function 
 $\psi({j}, \sigma)c^\dagger_{{j} \sigma}+\phi ({j},\pm\sigma)S^{\pm}_{{j} }$ ,   determines the spectrum of quasi-particle excitations  $\epsilon$ , the  $\psi({j},\sigma)$ and $ \phi(j, \sigma)$  amplitudes satisfy the following equations : 
\begin{eqnarray} 
&&\epsilon \psi({j}, \uparrow) -\lambda^*_{{j}} \psi({j},\downarrow)+\sum_{\textbf{1}}\psi({j+\textbf{1}}, \uparrow)=0, \nonumber\\
 &&\epsilon \psi({j},\downarrow ) -\lambda_{{j}}\psi({j},\uparrow)+ \sum_{\textbf{1}}\psi(j+\textbf{1},\downarrow )=0, \nonumber\\ 
&&\epsilon\phi({j},\uparrow) +\lambda^*_{{j}} \phi({j},\downarrow )=0,\nonumber\\ 
&&\epsilon \phi({j},\downarrow ) +\lambda_{j} \phi({j},\uparrow)=0, 
\end{eqnarray}
where sums over the nearest lattice sites. The variable $\lambda_{{j}}\to \pm \lambda_{\textbf{j}}$  is  identified with a static  ${Z}_2$- field defined at lattice sites.

The wave function describes singlet states of electrons and local moments at each lattice site. So the scattering of electron with spin flip in the $\lambda$-field occurs simultaneously with  spin flip of  local momentum at this lattice site.  Solution (4) conserves the total spin in the system, as  does the Hamiltonian  (1).

Let us consider the behavior of  electron liquid in a  static  ${Z}_2$-field in the case of free it configuration, when $\lambda_{{j}}=\lambda^*_{{j}} = \lambda$.   Two-particle wave function of  electrons  $\psi({j_1},\sigma_1;j_2,\sigma_2)  c^\dagger_{j_1,\sigma_1}c^\dagger_{j_2, \sigma_2}$ satisfies the  following equation  at $j_1\neq  j_2$
\begin{eqnarray} 
&&(\epsilon_1+ \epsilon_2)\psi(j_1,\sigma_1;j_2,\sigma_2) -\lambda\psi(j_1,-\sigma_1;j_2,\sigma_2)- \lambda\psi(j_1,\sigma_1;j_2,-\sigma_2)+ \nonumber\\&&
 \sum_{\textbf{1}}[\psi(j_1+\textbf{1},\sigma_1;j_2,\sigma_2) +
\psi(j_1,\sigma_1;j_2+\textbf{1},\sigma_2) ] =0,
\end{eqnarray}
at $j_1= j_2$ and arbitrary $S^z_j=\pm \frac{1}{2}$ 
\begin{eqnarray} 
&&(\epsilon_1+ \epsilon_2)\psi(j,\sigma;j,-\sigma) +\sum_{\textbf{1}}[\psi(j+\textbf{1},\sigma;j,-\sigma) +
\psi(j,\sigma;j+\textbf{1},-\sigma) ] =0,
\end{eqnarray} 
where $\epsilon_{1,2}=-2\cos {k}_{1,2}\pm\lambda$ are the energies of the electrons, $k_{1,2}$ are  wave vectors of the electrons. 

Due to the antiferromagnetic exchange interaction, local moments  and electrons form singlet states, electrons are scattered on local momets at each lattice site  at $j_1\neq j_2$ (see Eq (5)), for $j_1= j_2$ this scattering is forbidden (see Eq (6)). 
We use the Bethe ansatz to represent  the two-particle wave function $\psi({j_1},\sigma_1;j_2,\sigma_2) $:

$\psi({j_1},\sigma_1;j_2,\sigma_2) =A_{\sigma_1\sigma_2}(k_1 k_2)\exp (i k_1 j_1+i k_2j_2)- A_{\sigma_1\sigma_2}(k_2 k_1)\exp (i k_1 j_2+i k_2j_1)$ at $j_1<j_2$

$\psi({j_1},\sigma_1;j_2,\sigma_2) =A_{\sigma_2\sigma_1}(k_2 k_1)\exp (i k_1 j_1+i k_2j_2)- A_{\sigma_2\sigma_1}(k_1 k_2)\exp (i k_1 j_2+i k_2j_1)$  at $j_1>j_2$.

The A-amplitudes satisfy the continuity condition for the wave function at  $j_1=j_2$ and determine the two-particle scattering matrix ${\cal S}_{1,2}$.
The Bethe  function satisfies the following conditions $\psi({j},\sigma;j,-\sigma)+\psi({j},-\sigma;j,\sigma)=0$, $\sum_{\textbf{1}}[\psi(j+\textbf{1},\sigma;j,\sigma) +\psi(j,\sigma;j+\textbf{1},\sigma) ]=0$ for arbitrary  the A-amplitudes, as a result
$\psi({j},\sigma;j,\sigma)=0$ and  the Bethe function is a solution of  Eqs (5),(6).
This problem can be solved exactly in  the one dimension model so we calculate KI in the Kondo chain.

Solution for the two-particle scattering matrix  of electrons  ${\cal S}_{1,2}$ has the well-known form for the Hubbard chain  with on site interaction equal to $2\lambda$
 ${\cal S}_{1,2}=\frac{\sin k_1-\sin k_2 +i\lambda {\cal P}_{12}}{\sin k_1-\sin k_2 +i\lambda }$   \cite{LW}, here ${\cal P}_{12}$ is the spin-permutation operator.

The density of the ground state energy is equal  to
\begin{eqnarray} 
&
E=-\frac{2}{N}\sum_{k} \cos k - \lambda -\lambda +\frac{\lambda^2}{J},
\end{eqnarray}
where the first two terms taken into account the  energy of electrons, the  third term  takes into account the energy of  local moments. The solutions for the wave vectors satisfy the Bethe equations \cite{LW}. Taking into account the known relationship between the energies of the Hubbard chain with interactions of different sign for   half-filling
$E_0(-2|\lambda|)=-|\lambda|+E_0(2|\lambda|) $, where $E_0(2\lambda)$  value corresponds to first term to (7), we  can define that a solution $\lambda>0$ corresponds to a minimum of energy (7).
At half filling  the first term in (7) is equal to $E_0(2\lambda) =-4\int_0^\infty d\omega  \frac{J_0(\omega)  J_1 (\omega)}{\omega[1+\exp (\omega \lambda)]}$, here $J_{0,1}(\omega) $ are the Bessel functions of the first kind.

The problem is reduced to solution of the Hubbard chain at half filling.

\subsection*{The ground state}

The value of $\lambda$ satisfies the minimum of the ground state energy (7), it is solution of the following equation
\begin{eqnarray} 
&
2\frac{\lambda}{J}-2+\frac{d E_0(2\lambda) }{d \lambda}=0.
\end{eqnarray}

\begin{figure}[tp]
     \centering{\leavevmode}
\begin{minipage}[h]{.42\linewidth}
\center{
\includegraphics[width=\linewidth]{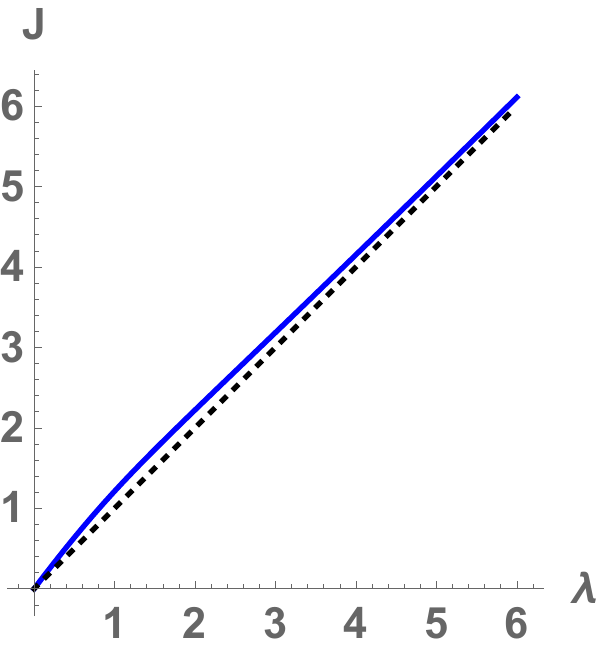} a)\\
                 }
   \end{minipage}
\begin{minipage}[h]{.33\linewidth}
\center{
\includegraphics[width=\linewidth]{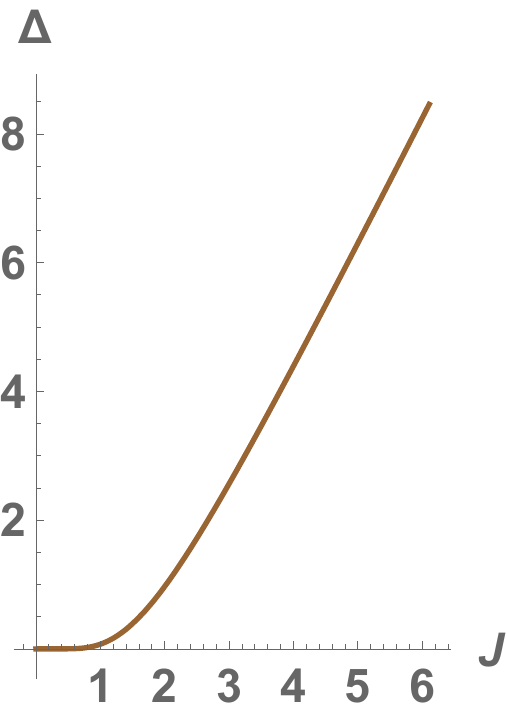} b)\\
                  }
 \end{minipage}
 \caption{
(Color online) The exchange integral $J$ as  function of $\lambda$ (dotted line $J=\lambda$ is shown for comparision)  a) , the gap in the spectrum of electrons $\Delta$ as  function of $J$  b).
 }
\label{fig:1}
\end{figure}

Numerical calculation of the exchange integtal  for different value of $\lambda$  is shown in Fig 1 a), the dotted line  $J=\lambda$ is a fairly good approximation.

Due to spin-charge separation that takes place in integrable quantum one-dimension models, two different types of elementary excitations are good determined: gapped spinless excitations carrying charge  and gapless, charge-neutral excitations carrying spin.
In the electron spectrum the gap \cite{LW}  opens at $\lambda\neq 0 $ and is equal to 
\begin{eqnarray} 
&
\Delta= -4+2\lambda+8\int_0^\infty d\omega \frac{ J_1 (\omega)}{\omega (1+\exp (\omega \lambda))}.
\end{eqnarray}

We can redefine the value of the gap \cite{LW,AA} and obtaine   $\Delta = \frac{8}{\lambda} \int_1^\infty dx \frac{\sqrt{x^2-1}}{\sinh(\pi x/\lambda)}$ ,   the asymptotic expressions at $\lambda \to 0$  $\Delta=8\sqrt{2 \lambda}/\pi \exp (-\pi/\lambda)$ 
and for large $\lambda$ $\Delta = 2\lambda-4+4\ln{2}/\lambda$. 
Let us visually illustrate the calculations of the dependence of the gap in the spectrum of quasi-particle excitations on the strength of the exchange interaction.  Numerical  calculation of gap  is presented in Fig 1 b).

\section*{The Anderson chain}
The Hamiltonian of the Anderson lattice is the sum of two terms, the first one determines the energies of $s$- and $d$-electrons and hybridization between them, the interaction term takes into account the on site repulsion of $d$-electrons  with different spins ${\cal H}_{A}={\cal H}_{s-d}+{\cal H}_{int}$
\begin{eqnarray}
&&{\cal H}_{s-d}= -
\sum_{<i,j>}\sum_{\sigma=\uparrow,\downarrow}
c^\dagger_{i,\sigma} c_{j,\sigma} +
v\sum_{j=1}^{N}\sum_{\sigma=\uparrow,\downarrow}
(c^\dagger_{j,\sigma} d_{j,\sigma}+ d^\dagger_{j,\sigma} c_{j,\sigma})+
(\epsilon_g+\frac{1}{2}U) \sum_{j=1}^{N}\sum_{\sigma=\uparrow,\downarrow}n_{j,\sigma},\nonumber\\&&
{\cal H}_{int}= U\sum_{j=1}^{N}\left( n_{j,\uparrow}-\frac{1}{2} \right) \left( n_{j,\downarrow}-\frac{1}{2} \right),
\label{eq:H}
\end{eqnarray}
where $c^\dagger_{j,\sigma},c_{j,\sigma}$ and $d^\dagger_{j,\sigma},d_{j,\sigma}$ are the fermion operators determined at a lattice site $j$, $U$ is the  value of the on site Hubbard interaction determined by the density operator $n_{j,\sigma}=d^\dagger_{j,\sigma}d_{j,\sigma}$,  $\epsilon_g$ is the energy of flat band of $d$-electrons,  $v$ determines the hybridization of $s$- and $d$-electrons.

We consider KI in the symmetric Anderson  chain at $\epsilon_g=-\frac{1}{2}U$ and  half filling.

\subsection*{Solution of the problem}

We redefine the term ${\cal H}_{int}$ of the Hamiltonian (10)  in the following form 
\begin{eqnarray} 
&& {\cal H}_{int}= -\frac{1}{2}U\sum_\textbf{j}(n_{{j},\uparrow}-n_{{j},\downarrow})^2.
\end{eqnarray} 

The Hubbard-Stratonovich transformation transforms  the  problem with interaction into a non-interacting one in a static  $\lambda$-field  \cite{K3}
   \begin{eqnarray}
&&{\cal H}_{eff}={\cal H}_0+2\sum_{{j}}\lambda_{{j}}(n_{{j},\uparrow}- n_{{j},\downarrow})+2 \sum_{{j}} \frac{\lambda^2_{{j}}}{U}.
   \end{eqnarray}

Let us consider equations for an  one-particle wave function 
 $\psi({j}, \sigma)c^\dagger_{{j} \sigma}+\phi (j,\sigma)d^\dagger_{{j} \sigma}$  which is solution of ${\cal H}_{eff}$ with energy $\epsilon$ : 
\begin{eqnarray} 
&&\epsilon \psi({j},\sigma) =v \phi({j},\sigma)-\sum_{\textbf{1}}\psi({j+\textbf{1}},\sigma), \nonumber\\
&&(\epsilon -2 \lambda_{{j}})\phi({j},\sigma ) =v\psi({j},\sigma), \nonumber\\ 
&&\epsilon \psi({j},-\sigma)=v \phi({j},-\sigma)-\sum_{\textbf{1}}\psi({j+\textbf{1}},-\sigma), \nonumber\\
&&(\epsilon +2 \lambda_{{j}})\phi({j},-\sigma ) =v\psi({j},-\sigma).
\end{eqnarray}
The variable $\lambda_{{j}}\to \pm \lambda_{\textbf{j}}$ determines the energies of $d$-electrons located at $j$-lattice site.
 We consider the solution of the problem for  free configuration of  a static  $\mathbb{Z}_2$- field at $\lambda_j=\lambda$.
Let us calculate low energy states of  $s$-electrons  with energies $|\epsilon|<<2|\lambda|$. In the $v\to 0$  limit $\lambda\to\frac{U}{4}$, two solutions 
$\epsilon=\epsilon_s (k)=-2\cos k$  and $\epsilon=\epsilon_d=\pm \frac{U}{2}$ correspond to  $s$- and $d$- noninteracting electrons. 
$s$-electrons hybridize with $d$-electrons with energy $-2|\lambda|$ located at the filled level, hybridization of $s$-electrons with $d$-electrons with energy $2|\lambda|$ corresponds to higher energy. In  a weak hybridization  $v<< \frac{U}{2}$
a correction of  $\epsilon_s$  near the Fermi energy shifts this energy  $\epsilon_s (k)=-2\cos k+\frac{v^2}{2|\lambda|}=-2\cos k  +\frac{2v^2}{U}$. 
The other limit $v> \frac{U}{2}$ corresponds to the intermediate-valence  $d$-states, in which there are no local moments at lattice sites.

Two-particle wave finction  $\psi({j_1},\sigma_1;j_2,\sigma_2)  c^\dagger_{j_1,\sigma_1}c^\dagger_{j_2, \sigma_2}$ of two electrons with energies $\epsilon_s (k_1) $  and  $\epsilon_s (k_2)$ satisfies the  following equations:

at $j_1\neq  j_2$
\begin{eqnarray} 
&&[\epsilon_s(k_1)+ \epsilon_s (k_2)]\psi(j_1,\sigma_1;j_2,\sigma_2) =\frac{v^2}{|\lambda|}]\psi(j_1,\sigma_1;j_2,\sigma_2)-
\sum_{\textbf{1}}[\psi(j_1+\textbf{1},\sigma_1;j_2,\sigma_2) +
\psi(j_1,\sigma_1;j_2+\textbf{1},\sigma_2) ],
\nonumber\\
\end{eqnarray}

at $j_1= j_2$  
\begin{eqnarray} 
&&[\epsilon_s(k_1)+ \epsilon_s (k_2)]\psi(j,\sigma;j,-\sigma)=-\sum_{\textbf{1}}[\psi(j+\textbf{1},\sigma;j,-\sigma) +
\psi(j,\sigma;j+\textbf{1},-\sigma) ] .
\nonumber\\
\end{eqnarray} 
At $j_1=j_2=j$  $s$-electrons with different spins hybridize with the first $d$-electron located at $j$ site with energy $2\lambda=-\frac{U}{2}$  and the  second $d$-electron located at $j$ site  with energy $-2\lambda=\frac{U}{2}$.  It leads to  an effective on-site attractive  interaction between $s$-electrons with value $-\frac{v^2}{|\lambda|}$ or  $-4\frac{v^2}{U}$. 
The behavior of the electron liquid in the Kondo chain and the symmetric Anderson chain is similar, the low-energy states of the Anderson chain are defined in the same way as in the Kondo chain when $2\lambda$ is replaced by $-4\frac{v^2}{U} $.

Iindirect interaction between electrons leads to attraction between them. The on site attraction between electrons allows us to offer an explanation of the nature of high-$T_c$ superconductivity.

\section*{Conclusion}

The  ground state  of electron liquid in the spin-$\frac{1}{2}$ Kondo and symmetric Anderson chains  is studied  at half-filling. 
In a static  ${Z}_2$- field electrons and  local moments  form local singlets
in the Kondo chain.  In free configuration of this field, the scattering of band electrons determines the spectrum of the quasi-particle excitations The scattering matrix of electrons coincides with that in the Hubbard chain. 
This makes it possible to calculate the gap in the quasi-particle spectrum  of KI for an arbitrary value of the interaction strength. 
In the limit of weak interaction, the results obtained coincide with those \cite{TS}.

In contrast to the Kondo lattice in the Anderson lattice, the on site indirect interaction between band electrons is attractive, which leads to a  hight-$T_c$ superconducting state  in three dimensional compounds. 

\section*{Appendix}
\begin{figure}[tp]
     \centering{\leavevmode}
\begin{minipage}[h]{.2\linewidth}
\center{
\includegraphics[width=\linewidth]{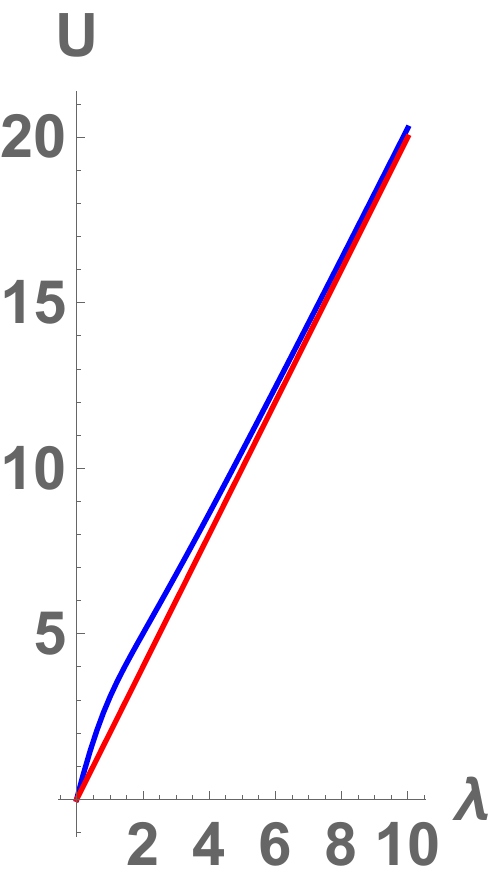} a)\\
                 }
   \end{minipage}
\begin{minipage}[h]{.35\linewidth}
\center{
\includegraphics[width=\linewidth]{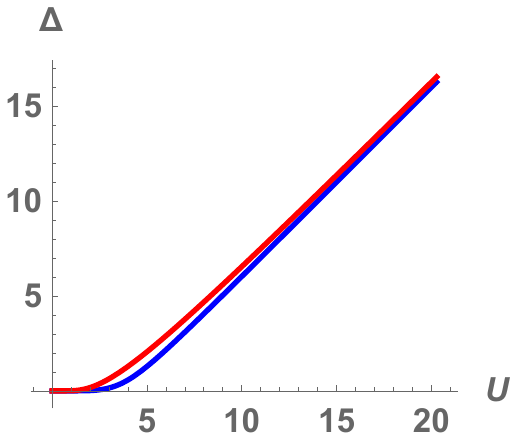} b)\\
                  }
 \end{minipage}
 \caption{
(Color online)  $U$  value as  function of $\lambda$ a): the calculation is represented by a blue line,  $U=2\lambda$ (red line) corresponds to exact solution. The gap in the spectrum of electrons $\Delta$ depending on $U$  b):  calculations by the mean field approach and the Bethe ansatz, the exact solution are marked with blue and red lines, respectively.
 }
\label{fig:2}
\end{figure}
Let's try to estimate the accuracy of the proposed calculation. As an example, we consider the solution of the Hubbard chain at half-filling and compare it with an exact result \cite{LW}. In the Hubbard Hamiltonian ${\cal H}_{Hub}={\cal H}_0+{\cal H}_{int}$, we redefine the term 
${\cal H}_{int}$ as 
\begin{eqnarray}
 &&{\cal H}_{int}= - U\sum_{j}\chi^\dagger_{j} \chi_j  -\frac{U}{2}N,
\end{eqnarray} 
where $\chi_j =c^\dagger_{j,\uparrow}c_{j,\downarrow}$.

The effective Hamiltonian, which follows from (16) and the Hubbard-Stratonovich transformation, is determined as 
\begin{eqnarray}
&&{\cal H}_{eff}={\cal H}_0+\sum_{j}(\lambda_{j}\chi_{j} + \lambda^*_{j}\chi^\dagger_{j})+\sum_{j} \frac{\lambda_{j}^2}{U}.
 \end{eqnarray}

Equations for one-particle wave function  $\psi({j}, \sigma)c^\dagger_{j \sigma}$ coincide  with (4) and for two-particle wave function $\psi({j_1},\sigma_1;j_2,\sigma_2)  c^\dagger_{j_1,\sigma_1}c^\dagger_{j_2, \sigma_2}$ with (5),(6). 
In the case of a free configuration of $Z_2$-field, equaiion for $\lambda$  follows from the  minimum of the density 
of the ground state energy
$E=-\frac{2}{N}\sum_{k} \cos k - \lambda  +\frac{\lambda^2}{U}$.
Numerical calculation $U$ as function of $\lambda$  and comparison  with $U=2 \lambda$ in \cite{LW}  are shown in Fig 2a), where red  and blue lines correspond to exact solution and calculation proposed.  In Fig 2b) the calculations of the gap are also presented. 

Sufficiently good agreement between the calculation  and the exact solution allows us to solve the problem for the case of an arbitrary exchange interaction.

\section*{Acknowledgments} The author thanks the Bar Ilian University  and personally Prof. R. Berkovits  and D.Golosov for support.

 \section*{Author contributions statement} I.K. is an author of the manuscript 

\section*{Additional information} The author declares no competing financial interests. \\ 

\section*{Availability of Data and Materials} All data generated or analysed during this study are included in this published article.\\ Correspondence and requests for materials should be addressed to I.K. 

\end{document}